# The fourth microlensing planet revisited

**Philip Yock**                                                    **History**

The fourth microlensing planet, otherwise known as OGLE-2005-BLG-169Lb, was discovered by a collaboration of US, NZ, Polish and UK astronomers in 2005-2006. Recently the results were confirmed by the Hubble Space Telescope and by the Keck Observatory. OGLE-2005-BLG-169Lb is the first microlensing planet to receive such confirmation. Its discovery and confirmation are described here in an historical context.

### Introduction

In 1936 an engineer of Moravian extraction with an interest in science, Rudi Mandl, visited Einstein at Princeton University and requested Einstein to publish his belief that the bending of light by the gravitational field of a star was comparable to the well-known physics of optical lenses, and that light from a distant star in an eclipse could be greatly brightened by the nearer star[1]. After some discussion Einstein agreed to Mandl's request and he published a short paper[2] in *Science* on the "lens-like action of a star by the deviation of light in the gravitational field". This short paper arguably spawned one of New Zealand's more productive contributions to astronomy some 60 years later.

This article describes one item amongst those contributions, the discovery of a Neptune-like planet halfway to the centre of the Milky Way that was made ten years ago. The discovery was met with scepticism at the University of Auckland, but it was confirmed recently by the Hubble Space Telescope and by the Keck Observatory.

According to *Science News*, Mandl was born in Moravia, served for Austria in WWI, was captured and sent to Siberia, escaped, completed a degree in electrical engineering, moved to South America, Germany and thence to the US where in the 1930s he pursued his interest in science[1]. This culminated in his meeting with Einstein at Princeton and the subsequent publication of the above paper in *Science*. A recently released photograph of Mandl appears below.

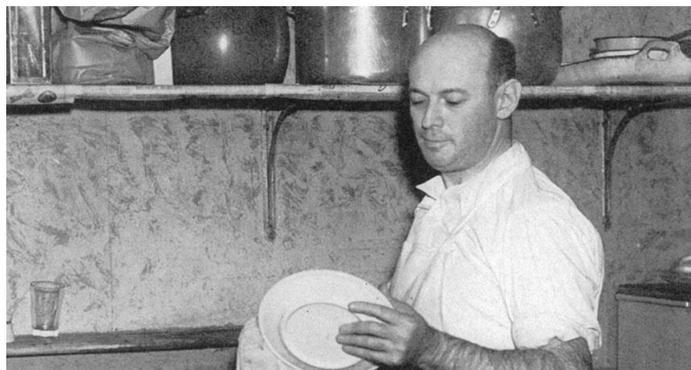

**Figure 1.** Rudi Mandl, shown here in the 1930s in Washington DC where he made a living as a busboy and washing dishes, persuaded Einstein to publish his 1936 paper on gravitational lensing. Photo courtesy Tom Siegfried of *Science News*.

Einstein thought there was no great chance of observing the lensing effect, but he did not foresee the power of the modern wide-field telescope equipped with large CCD detectors that can accurately monitor the brightnesses of millions of stars simultaneously. These enable the observation of the bright eclipses envisaged by Mandl in 1936, and also the discovery of distant planets as was foreseen by Sidney Liebes in 1964[3].

The study of planets orbiting stars other than the Sun blossomed in the last 20 years following the discovery by Swiss astronomers Michel Mayor and Didier Queloz[4] of the first planet orbiting a sun-like star in 1995. Now over 1,500 exoplanets have been found using various techniques.

New Zealand pulled its weight in this enterprise. A group led by Professor John Hearnshaw at the University of Canterbury was one of the first to seek sub-stellar companions to stars[5]. Currently they are monitoring our nearest neighbour stars alpha Centauri A and B for evidence of small, tell-tale wobbles indicative of orbiting Earth-like planets[6].

As alluded to above, NZ astronomers also used the Mandl-Einstein-Liebes lens-like action of stars to find exoplanets[7,8]. The technique, which is known today as "gravitational microlensing", is illustrated below where a red dwarf is shown lensing a more distant solar-like star, leading to the formation of an Einstein ring around the red dwarf. Even though the ring is generally too small to be resolved, the distant star appears to be brightened or magnified because the surface area of the ring is larger than the area of the distant star projected to the position of the lens.

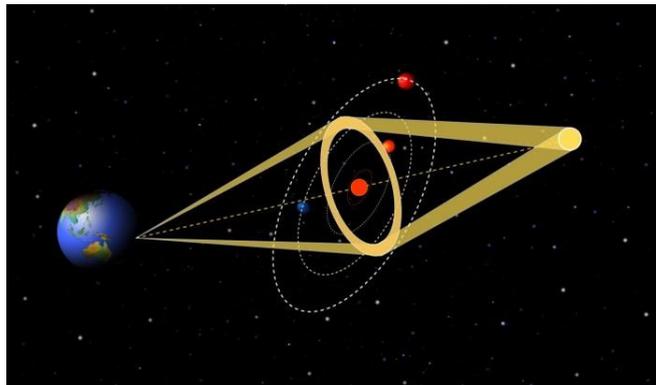

**Figure 2.** Formation of an Einstein ring by the gravitational field of a star. The example shows a yellow solar-like source star being lensed by an aligned red dwarf with orbiting planets.

The microlensing effect is most easily witnessed in the southern sky in the dense stellar fields towards the centre of the Galaxy. Hence the presence of microlensing telescopes in NZ, Australia, South Africa and Chile. The largest of these is the MOA telescope at the Mt John University Observatory in Canterbury. This is operated by the Japan/NZ MOA group[9] and is shown below. A Polish group known as OGLE[10] operates a comparable telescope in Chile.

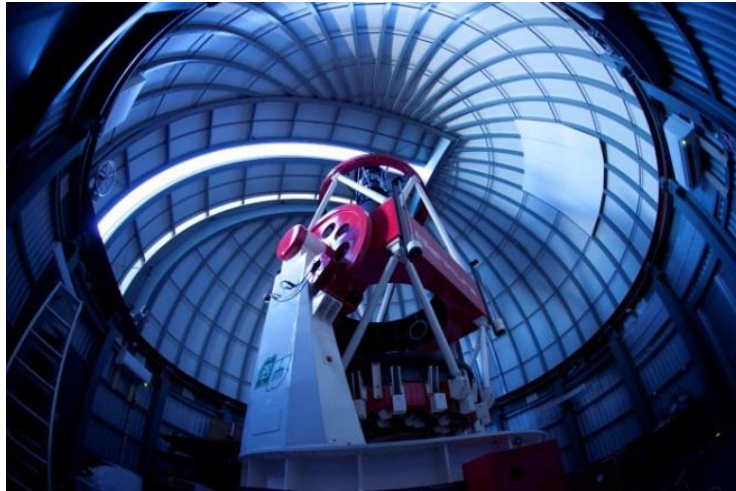

**Figure 3.** The 1.8m MOA telescope at Mt John was funded and built primarily by Japan to a NZ-supplied design[7]. Photo courtesy Maki Yanagimachi of Earth & Sky Ltd, Tekapo.

## Planet detection by gravitational microlensing at high magnification

Although the OGLE and MOA groups commenced their observations in the early 1990s the first microlensing planet was not discovered until 2004[11]. The first ten years can be described as a familiarisation period when the observational and analytical techniques of microlensing were honed.

One of the main developments of the early years was the realization that exoplanets could be detected with the telescopes then in use in microlensing events of high magnification. The reasoning went as follows.

It is first noted that an alignment such as that depicted in Fig. 2 can only last temporarily because all stars in the Galaxy are in motion. As the lens and source stars move into alignment, and out of it, the magnification grows and wanes over a period of some 20 days.

The maximum magnification that can be attained occurs at perfect alignment. If a solar like star in the galactic bulge is lensed by a red dwarf in the galactic disc, which is the most commonly observed situation, an elementary calculation shows that the radius of the Einstein ring $\sim 2$ AU, and that the maximum magnification is $\sim 2000$, or about 8 magnitudes. Mandl was thus correct, an eclipsed star can be greatly magnified. In nearly all cases, however, the two stars never achievement perfect alignment, and the maximum magnification is < 2000.

If the lens star has a planetary system orbiting it that resembles our solar system, planets are likely to be in the vicinity of the Einstein ring and, although much less massive than a star, they can perturb the ring appreciably and hence betray their presence.

Unlikely as it may seem, this provides a sensitive means for detecting planets, because it is a resonant process. The lens star magnifies the source star, and a planet near the ring perturbs the magnified image. This favours the detection of planets in microlensing events of high magnification. Several papers were published on this around the year 2000, including some by an Auckland group consisting of Drs Nicholas Rattenbury, Ian Bond and myself[12-15].

Two of the first four microlensing planets were found in events of high magnification, and, of these, the last one was found in the microlensing event OGLE-2005-BLG-169 which reached a maximum magnification of 800. The closest alignment in this event thus came close (actually within 1 μas) to the perfect alignment envisaged by Mandl, and this resulted in the discovery of a Neptune-like planet that was confirmed recently by the Hubble Space Telescope and the Keck Observatory. The following remarks describe in general terms the initial observations of this planet and the more recent confirmatory observations.

### The fourth microlensing planet, OGLE-2005-BLG-169Lb

OGLE-2005-BLG-169Lb was found in the 169[th] lensing event found by the OGLE group in the galactic bulge in 2005. The planet orbits the lens star of the event, hence the designation Lb. A plot of the observed magnification versus time is shown in Fig. 4.

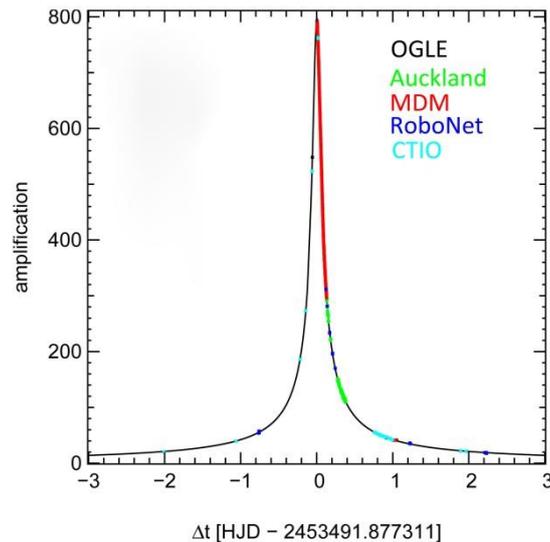

**Figure 4.** Light curve of OGLE-2005-BLG-169 as observed by the 1.3m OGLE telescope in Chile, the 40cm telescope at the Auckland Observatory, the 2.4m MDM telescope at Kitt Peak, Arizona, the 2m RoboNet telescope at Hawaii and the 1.3m SMARTS telescope at CTIO in Chile.

No data were obtained by the MOA telescope on this event because unfortunately it fell in one of the small gaps between CCDs on the 80M pixel MOA camera. However, following our interest at Auckland in high magnification events, I enquired of Professor Andrew Gould of Ohio State University (OSU), who had co-ordinated the observations of OGLE-2005-BLG-169, if he might make the data available to us on the event. This he kindly agreed to do.

Visual examination of the light curve revealed slight deviations, $\sim \pm 1\%$, in the MDM data from the smooth light curve expected for a planetless lens. Though too small to be seen on Fig. 4, they appeared to be possibly indicative of a low-mass planet, and they piqued our interest.

We therefore decided to analyse the data fully at Auckland, and the Ohio group followed suit. To ensure that the best possible fit to the data was obtained, systematic and independent searches for all possible fits were commenced at the Tamaki Campus of the University of Auckland and at OSU. The Tamaki search was carried out by Dr Christine Botzler, Mr Stephen Swaving (initial work only) and myself using code that had previously been written by Dr Lydia Philpott for just such a purpose.

Both the Tamaki and OSU groups used a ray shooting method to simulate the gravitational lens. This was similar to the technique commonly used by designers of optical lenses. Rays were shot through a gravitational lens with each ray suffering the deflection $4GM/c^2b$ predicted by general relativity for each component of the lens. Here *G* is Newton's constant of gravitation, *M* the mass of a lens component, *c* the speed of light and *b* the impact parameter of a ray with respect to the mass *M*. In this way all possible lens configurations were tested.

Both groups found two possible solutions with almost equal chi squares and likelihoods. They had *q* ≈ 8 × 10⁻⁵ and *α* ≈ 118° or *q* ≈ 6 × 10⁻⁵ and *α* ≈ 89°. Here *q* denotes the planet:star mass ratio of the lens, and *α* the angle between the planet:lens and source:lens axes. The Tamaki results are shown below.

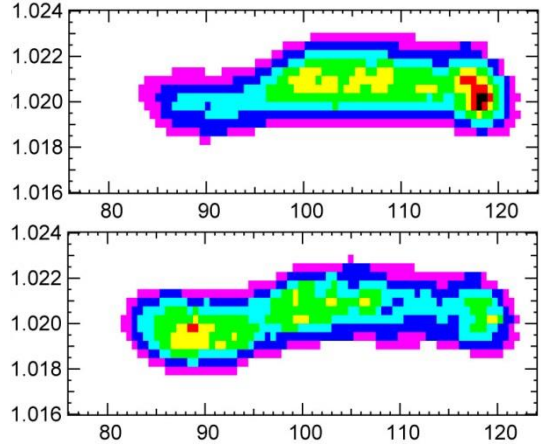

**Figure 5.** Likelihood maps of OGLE-2005-BLG-169Lb found at Tamaki in 2006 for planet:star mass ratios of 8 × 10⁻⁵ (upper) and 6 × 10⁻⁵ (lower). The horizontal axis denotes the position angle *α* of the planet, and the vertical axis its projected distance from the host star in units of the Einstein radius $r_E$. Black denotes regions of highest probability within 1 *σ*, red 2 *σ*, etc.

Detailed measurements of the microlensing light curve yielded the additional results that *μ* = 7 − 10 mas yr⁻¹ where *μ* denotes the relative proper motion of the lens and source stars, and that $\vartheta_E$ ≈ 1.0 mas where $\vartheta_E$ is the angular Einstein radius of the lens[16]. These results were obtained from the planetary deviation to the light curve. The lens star was, however, not indentifiable from the 2005 data as it was superimposed almost perfectly on the source star at the time.

By making the reasonable assumption that the lens star was a red dwarf with mass ∼ 0.4 M$_\odot$ at a distance about midway to the galactic bulge, i.e. at a distance ∼ 4 kpc, it followed that the planet had a mass ∼ 0.6 M$_{Neptune}$ and that it was located at a projected separation of ∼ 2.5 AU from its host star, i.e. well beyond the snowline of the host star. This suggested a Neptune-like identification of OGLE-2005-BLG-169Lb. Comparison with all the data that were available at the time indicated that such planets must be common members of planetary systems in the Galaxy[16].

The above results were important at the time and they remain important today. Neptune-like planets have orbital periods of order some years. This renders their detection by either the radial velocity or transit techniques very difficult. An artist's impression of OGLE-2005-BLG-169L is reproduced below.

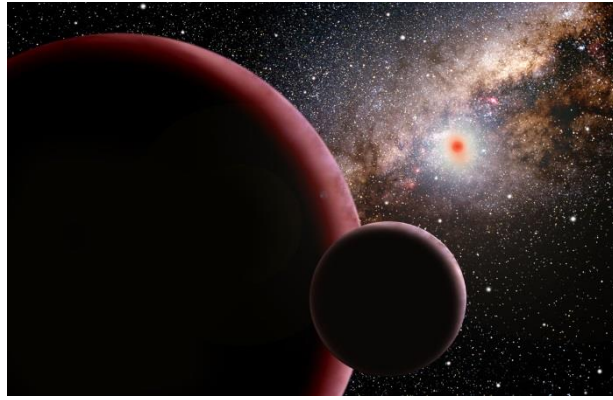

**Figure 6.** Artist's image of OGLE-2005-BLG-169L showing the orange K5 lens star half way to the centre of the Galaxy being orbited by a Neptune-like planet. Artistic license has been invoked to add a moon to the planet. Courtesy David Aguilar, Center for Astrophysics, Harvard University.

**Confirmation of planet OGLE-2005-BLG-169Lb by the Hubble Space Telescope and the Keck Observatory**

Concern was expressed[17,18] at the University of Auckland about the research practices described above. Certainly they were new, but they were planned and executed with care, and they received independent confirmation as noted above.

Further checks subsequently became possible and these have now been successfully conducted and reported[19,20]. They entailed observing the divergence of the source and lens stars from one another since 2005. The feasibility of doing this had been noted previously[13]. As noted above, the rate of divergence of the stars was predicted to be 7-10 mas yr$^{-1}$ from the planetary fit to the microlensing data. This prediction was tested by merely waiting until an observable separation had accrued.

The angular resolution of the Hubble Space Telescope (HST) is 1.22 $\lambda/D$ according to the Rayleigh criterion where $D$ denotes the diameter of the telescope and $\lambda$ denotes wavelength. The diameter of the HST is 2.4m, so for the HST broadband filter centred on 814 nm the Rayleigh criterion predicts a resolution of 85 mas. However, the actual resolution achieved with a CCD detector is somewhat better than this, as the shape or "point spread function" of the stellar image is sampled. The actual resolution of the HST is approximately 50 mas. This implies the lens and source stars in OGLE-2005-BLG-169 should have been resolvable from about 2011 onwards.

Figure 7 below shows an image of the lens and source stars 6.5 years after the microlensing event that was taken with the 814 nm HST filter. The stars are resolved, as expected. The measured separation is 49 mas. This corresponds to a divergence velocity of 7.5 mas yr$^{-1}$ which is consistent with the expectation. Figure 8 shows a similar image taken 8.2 years after the event with the Keck Observatory using adaptive optics. It also corresponds to a divergence rate of 7.5 mas yr$^{-1}$.

OGLE-2005-BLG-169Lb was the first microlensing planet to receive this type of confirmation. The divergence of the lens and the source stars after a planetary microlensing event had been observed before[7], but not with sufficient precision to definitively confirm the planetary model.

The above observations leave no doubt that a Neptune-like planet was detected in the microlensing event OGLE-2005-BLG169. However, they do more. The observations of the lens star by the HST and Keck telescopes in various passbands allow the lens star to be identified and its distance to be determined[13]. It was found to be a K5 type main sequence star of mass 0.65 ± 0.05 M$_\odot$ at a distance of 4.0 ± 0.4 kpc[19,20]. These results imply that the planet has a mass of 0.85 ± 0.08 M_{Neptune}, that its projected separation at the time of the event was 3.4 ± 0.3 AU, and also that lower solution in Fig. 5 with $q \approx 6 \times 10^{-5}$ and $\alpha \approx 89°$ was the preferred one. Further details are given in references 19 and 20.

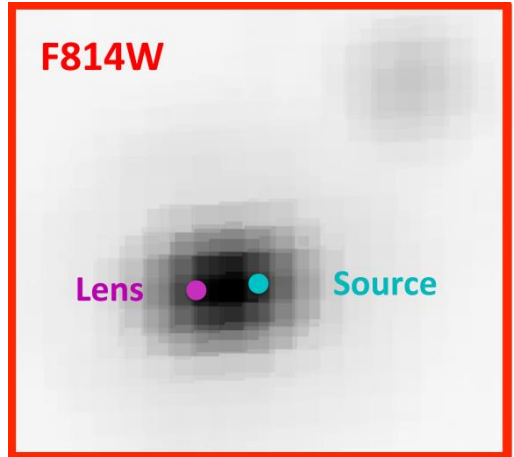

**Figure 7.** Hubble Space Telescope image at 814 nm measuring 0.42″ × 0.37″ showing the lens and source stars of OGLE-2005-BLG-169 diverging from one another 6.5 years after the microlensing event[19].

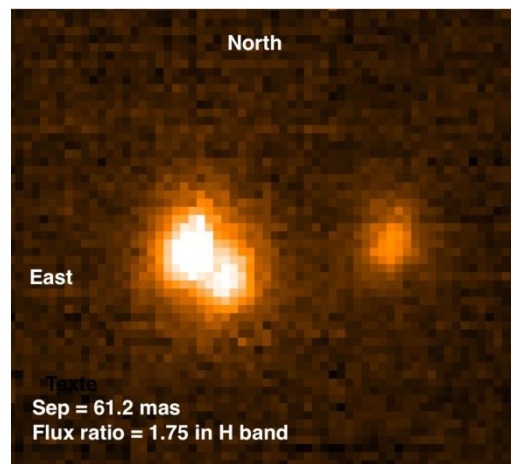

**Figure 8.** Keck Observatory image of the lens and source stars of OGLE-2005-BLG-169 diverging from one another 8.2 years after the event[20]. To facilitate comparison with Fig. 7, the above figure needs to be rotated approximately 40° anticlockwise. The brightest star is the lens star, and the second brightest the source star. The dimmest star is unrelated.

## Concluding remarks

Hopefully the above example has clarified the main features of planet detection by the gravitational microlensing method. The physics behind the process is fully contained in Einstein's simple formula for the bending of a ray of light in the gravitational field, *4GM/c²b*, but this yields a remarkable diversity of phenomena.

The present example of OGLE-2005-BLG-169Lb provided a relatively straightforward example in which a K5 main sequence star half way to the centre of the Galaxy passed almost immediately in front of (within 1 µas) of a main sequence star in the central bulge of the Galaxy that led to the formation of an almost complete Einstein ring with angular radius 1 mas. The K5 star hosted a Neptune-like planet which was (in projection) close to the ring. It caused a small but detectable perturbation on the lensing light curve.

Other microlensing events have been found where a lens star with two planets is required to explain the data, or where the motion of the Earth around the sun needs to be allowed for, or the motion of the source star about a companion star, or rotation of the lens during an event, or even where the different locations of telescopes on the Earth's surface need to be taken account of. Triple lenses consisting of a star and two planets are the most complicated found to date. Single lenses consisting of an isolated object of planetary mass have also been detected. Publications on all the above appear at the MOA website[9]. In general, one appeals to Occam's razor to select the simplest geometry that can accommodate the data for any event.

New microlensing projects have recently commenced operation, and others are on the drawing boards. A US group known as LCOGT[21] recently installed eight robotic 1m telescopes in Australia, South Africa and Chile to monitor microlensing events round the clock, a Korean group known as KMTNet[22] recently installed three wide-field 1.6m telescopes at the same locations with a similar motivation, and NASA has plans for a 2.4m wide-field space telescope known as WFIRST[23] to be launched in about ten years. All these projects will study exoplanets between us and the centre of the Galaxy.

The author's personal dream is to see some NZ astronomers migrate southwards to the pristine skies and long winter nights of Antartica, whilst not abandoning the clear skies of Mt John. Antarctica may well be the world's premier site for detecting planets by the transit technique, and also for astronomy at terahertz frequencies[24]. It is certainly the case that observations of planetary transits have led to dramatic advances in our knowledge of exoplanets in recent years. The recent detection, for example, of an 11.2 billion year old star 117 light years away hosting five terrestrial exoplanets raises dramatic and far-reaching questions[25].

Finding new planets can only be described as discovery science, especially for those planets that are without analogues in our solar system. In this regard, I concur with the recommendation[26] of the Science Adviser to the Prime Minister when he wrote recently that "while NZ is a modest component of the international research effort … our contribution to the global effort in discovery science should be protected". It is to be hoped that future explorations of planets in our Galaxy are protected.

## Acknowledgments

It is a pleasure to thank colleagues, amateur and professional, and students and friends from several countries who assisted the above described research. Financial assistance by the NZ Marsden Fund for most of the last 20 years is also acknowledged.

---

2/42 Reihana Street,
Orakei,
Auckland,
New Zealand.
p.yock@xtra.co.nz